\documentclass[a4paper,12pt,eqno]{article}
\usepackage{theorem}
\usepackage{latexsym,amssymb,amsfonts,amsmath}
\usepackage{graphicx}
\newtheorem{thm}{Theorem}[section]
\newtheorem{lem}[thm]{Lemma}

\newtheorem{pro}[thm]{Proposition}

\theoremheaderfont{\scshape}
\newcommand{\RM}{\mathbb{R}}
\newcommand{\ZM}{\mathbb{Z}}

\newcommand{\CM}{\mathbb{C}}

\newcommand{\ket}[1]{|#1\rangle}

\title{{\Large {\bf The uniform measure for discrete-time quantum walks \\ in one dimension}}}
\author{
{\small Norio Konno}\\
{\scriptsize Department of Applied Mathematics, 
Faculty of Engineering, 
Yokohama National University}\\
{\scriptsize Hodogaya, Yokohama 240-8501, Japan}\\
{\scriptsize e-mail: konno@ynu.ac.jp}\\
}
\date{\empty }
\begin{document}
\maketitle

\par\noindent
\begin{small}
\par\noindent
{\bf Abstract}. We obtain the uniform measure as a stationary measure of the one-dimensional discrete-time quantum walks by solving the corresponding eigenvalue problem. As an application, the uniform probability measure on a finite interval at a time can be given.

\footnote[0]{
{\it Abbr. title:} Uniform measure for quantum walks
}
\end{small}

\setcounter{equation}{0}
\section{Introduction \label{intro}}
The quantum walk (QW) has attracted much attention as a quantum analog of the classical random walk for the last decade. There are two types of QWs. One is the discrete-time walk and the other is the continuous-time one. In this paper, we focus on the discrete-time case. The QW on $\ZM$ was intensively studied by Ambainis et al. \cite{AmbainisEtAl2001}, where $\ZM$ is the set of integers. A number of non-classical properties of the QW have been shown, for example, ballistic spreading, anti-bellshaped limit density, localization. The review and books on QWs are Kempe \cite{Kempe2003}, Kendon \cite{Kendon2007}, Venegas-Andraca \cite{VAndraca2008, Venegas2013}, Konno \cite{Konno2008b}, Cantero et al. \cite{CanteroEtAl2013}, Manouchehri and Wang \cite{MW2013}.

One of the interesting problems is to find a way to produce the uniform measure using a QW. This paper presents the way in which we obtain the uniform measure as a stationary measure for the space-homogeneous QW on $\ZM$ by solving the corresponding eigenvalue problem. Moreover we consider the relation between stationary measure, time-averaged limit measure, and weak limit measure for our QWs.

We will explain some related works. Konno et al. \cite{KLS2013} computed time-averaged limit measure for $X_n$ and proved weak limit theorem for $X_n/n$, where $X_n$ is a two-state QW at time $n$ on $\ZM$ with one defect at the origin. Remark that the QW is space-inhomogeneous. Moreover they obtained the stationary measure which has exponential decay with respect to the position. Therefore it is not the uniform measure. However, in a special space-homogeneous case (the Hadamard walk), the stationary measure becomes the uniform measure. In fact, our result is a generalization of this Hadamard walk case. Furthermore they showed that a stationary measure with exponential decay for the QW starting from infinite sites is identical to a time-averaged limit measure for the same QW starting from just the origin. Konno and Watanabe \cite{KW2013} calculated a stationary measure of another type of two-state space-inhomogeneous QW on $\ZM$ with one defect at the origin which was introduced and studied by W\'ojcik et al. \cite{WojcikEtAl2004}. This stationary measure also has exponential decay and is not the uniform measure. Machida \cite{Machida2013} proved that $X_n/n$ converges weakly to a uniform probability measure on $[-|\cos (\theta)|, |\cos (\theta)|]$, where $X_n$ is the two-state space-homogeneous QW on $\ZM$ determined by $U(\theta)$ (given by Eq.(\ref{akina})) starting from suitable infinite sites. Our method is related to the CGMV method \cite{CanteroEtAl2013}. In fact, both methods deal with the eigenvalue problem. However, our approach is mainly based on generating functions and CGMV is the use of the CMV matrices combined with the theory of orthogonal polynomials on the unit circle. To clarify the relation between them would be an interesting problem.

Under this background, we prove that any uniform measure belongs to the set of stationary measures for the two-state space-homogeneous QW on $\ZM$. Moreover, we show that the same statement holds for the three-state space-homogeneous Grover walk on $\ZM$. We should remark that as an obvious case (the eigenvalue is 1 for the eigenvalue problem), this is valid for general $N$-state Grover walk also.

From now on we give a detailed explanation of our result. In this paper, we mainly consider the two-state QW. So we focus on this case, however the similar argument holds for the general $N$-state QW.

The two-state QW on $\ZM$ is determined by a $2 \times  2$ unitary matrix $U$. Let ${\cal M}_s = {\cal M}_s (U)$ be the set of stationary measures of the QW (the precise definition is given in the next section). For any $c>0$, $\mu_{u, \ZM}^{(c)}$ denotes the uniform measure with parameter $c$, i.e., 
\begin{align*}
\mu_{u, \ZM}^{(c)} (x) = c \qquad (x \in \ZM). 
\end{align*}
In a similar way, for any $\psi (\not=0) \in \CM$, $\Psi_{u, \ZM}^{(\psi)}$ denotes the uniform amplitude with parameter $\psi$, i.e., 
\begin{align*}
\Psi_{u, \ZM}^{(\psi)} (x) = \psi \begin{bmatrix} 1 \\ 1 \end{bmatrix}  \qquad (x \in \ZM). 
\end{align*}
Moreover, for any $M \in \{1,2, \ldots \}$, $\mu_{u, [-M,M]}$ denotes the uniform probability measure on $\{-M,-M+1, \ldots , M \}$, i.e., 
\begin{align*}
\mu_{u, [-M,M]} (x) = \frac{1}{2M+1} \qquad (x \in \{-M,-M+1, \ldots , M \}).  
\end{align*}
In addition, for any $c>0$ and $a < b$, $\mu_{u, (a,b)}^{(c)}$ denotes the uniform measure on $(a,b)$, i.e., 
\begin{align*}
\mu_{u, (a,b)}^{(c)} (x) = c \qquad (x \in (a,b)).
\end{align*}

Then our main result is that for any $c>0$, there exists an initial state such that
\begin{align*}
\mu_{u, \ZM}^{(c)} \in {\cal M}_s (U).
\end{align*}
Remark that the same conclusion holds for the three-state Grover walk (see Sect. 6). To obtain the result, we solve the corresponding eigenvalue problem:
\begin{align*}
U^{(s)} \Psi^{(\lambda)} = \lambda \Psi^{(\lambda)},
\end{align*}
where $U^{(s)}$ is $\infty \times \infty$ unitary matrix which determines the evolution of the QW. Here the eigenvalue $\lambda \in \CM$ satisfies $|\lambda|=1$.

As an application of the result, we easily see that if we take the following state;
\begin{align*} 
\Psi_0 (x)
=
\begin{cases} 
\Psi^{(\lambda)} (x) & \text{$(|x| \leq 2M)$, }\\ 
\\
\begin{bmatrix} 0 \\ 0 \end{bmatrix} & \text{$(|x| \geq 2M+1)$,} 
\end{cases} 
\end{align*} 
for $M \in \{1,2, \ldots \}$, as an initial state of the QW, then the measure at time $M$, $\mu_M$, becomes the uniform probability measure on $\{-M,-M+1, \ldots , M \}$. That is, 
\begin{align*} 
\mu_M (x) = \mu_{u, [-M,M]} (x) \qquad (|x| \le M). 
\end{align*} 
This fact could be useful for quantum information processing. 

We should remark that for the corresponding classical random walk in which the walker moves one step to the left with probability $p$ and to the right with probability $q$ with $p+q=1 \> (p,q \in [0,1])$, it is known that the uniform measure $\mu_{u, \ZM}^{(c)} \> (c>0)$ is the stationary measure.

The rest of the paper is organized as follows. Section \ref{def} gives the definition of our model. Sections \ref{azero}, \ref{bzero}, and \ref{abcdzero} deal with two-state QWs with $a=0, \> b=0$, and $abcd \not=0$ cases respectively. In Sect. \ref{gw}, we consider three-state Grover walk case. Section \ref{sum} is devoted to summary.

\section{Definition of Two-State Model \label{def}}
This paper mainly treats the two-state QW on $\ZM$. In fact, we deal with two-state case except Sect. \ref{gw}. So this section gives the definition of the two-state QW. Section \ref{gw} deals with three-state case, so we will present the definition of the model there.

The discrete-time QW is a quantum version of the classical random walk with additional degree of freedom called chirality. The chirality takes values left and right, and it means the direction of the motion of the walker. At each time step, if the walker has the left chirality, it moves one step to the left, and if it has the right chirality, it moves one step to the right. Let define
\begin{align*}
\ket{L} = 
\left[
\begin{array}{cc}
1 \\
0  
\end{array}
\right],
\qquad
\ket{R} = 
\left[
\begin{array}{cc}
0 \\
1  
\end{array}
\right],
\end{align*}
where $L$ and $R$ refer to the left and right chirality states, respectively.  

The time evolution of the walk is determined by a $2 \times 2$ unitary matrix $U$, where
\begin{align*}
U =
\left[
\begin{array}{cc}
a & b \\
c & d
\end{array}
\right],
\end{align*}
with $a, b, c, d \in \mathbb C$ and $\CM$ is the set of complex numbers. To define the dynamics of our model, we divide $U$ into two matrices:
\begin{eqnarray*}
P =
\left[
\begin{array}{cc}
a & b \\
0 & 0 
\end{array}
\right], 
\quad
Q =
\left[
\begin{array}{cc}
0 & 0 \\
c & d 
\end{array}
\right],
\end{eqnarray*}
with $U =P+Q$. The important point is that $P$ (resp. $Q$) represents that the walker moves to the left (resp. right) at any position at each time step.

One of the typical class considered here is 
\begin{align}
U = U (\theta) = 
\left[
\begin{array}{cc}
\cos \theta & \sin \theta \\
\sin \theta & - \cos \theta 
\end{array}
\right], 
\label{akina}
\end{align}
where $\theta \in [0, 2 \pi)$. If $\theta = \pi/4$, then this model is equivalent to the {\it Hadamard walk} determined by the Hadamard matrix $H$:
\begin{align*}
H=\frac{1}{\sqrt2}
\left[
\begin{array}{cc}
1 & 1 \\
1 &-1 
\end{array}
\right].
\end{align*}
Let $\Psi_n$ denote the amplitude at time $n$ of the QW:  
\begin{align*}
\Psi_{n}
&= {}^T\![\cdots,\Psi_{n}^{L}(-1),\Psi_{n}^{R}(-1),\Psi_{n}^{L}(0),\Psi_{n}^{R}(0),\Psi_{n}^{L}(1),\Psi_{n}^{R}(1),\cdots ],
\\
&= {}^T\!\left[\cdots,\begin{bmatrix}
\Psi_{n}^{L}(-1)\\
\Psi_{n}^{R}(-1)\end{bmatrix},\begin{bmatrix}
\Psi_{n}^{L}(0)\\
\Psi_{n}^{R}(0)\end{bmatrix},\begin{bmatrix}
\Psi_{n}^{L}(1)\\
\Psi_{n}^{R}(1)\end{bmatrix},\cdots\right],
\end{align*}
where $T$ means the transposed operation. Then the time evolution of the walk is defined by 
\begin{align*}
\Psi_{n+1}(x)= P \Psi_{n} (x+1) +  Q \Psi_{n}(x-1).
\end{align*}
That is 
\begin{align*}
\begin{bmatrix}
\Psi_{n+1}^{L}(x)\\
\Psi_{n+1}^{R}(x)
\end{bmatrix}
=
\begin{bmatrix}           
a \Psi_{n}^{L}(x+1)+b \Psi_{n}^{R}(x+1)\\
c \Psi_{n}^{L}(x-1)+d \Psi_{n}^{R}(x-1)
\end{bmatrix}.
\end{align*}
Now let
\begin{align*}
U^{(s)}=\begin{bmatrix}
\ddots&\vdots&\vdots&\vdots&\vdots&\cdots\\
\cdots&O&P&O&O&O\cdots\\
\cdots&Q&O&P&O&O\cdots\\
\cdots&O&Q&O&P&O\cdots\\
\cdots&O&O&Q&O&P\cdots\\
\cdots&O&O&O&Q&O\cdots\\
\cdots&\vdots&\vdots&\vdots&\vdots&\ddots
\end{bmatrix}\;\;\;
with\;\;\;O=\begin{bmatrix}0&0\\0&0\end{bmatrix}.
\end{align*}
Then the state of the QW at time $n$ is given by
\begin{align*}
\Psi_{n}=(U^{(s)})^{n}\Psi_{0},
\end{align*} 
for any $n\geq0$. Let $\mathbb{R}_{+}=[0,\infty)$. Here we introduce a map 
$\phi:(\mathbb{C}^{2})^{\mathbb{Z}}\rightarrow \mathbb{R}_{+}^{\mathbb{Z}}$
such that if
\begin{align*}
\Psi= {}^T\!\left[\cdots,\begin{bmatrix}
\Psi^{L}(-1)\\
\Psi^{R}(-1)\end{bmatrix},\begin{bmatrix}
\Psi^{L}(0)\\
\Psi^{R}(0)\end{bmatrix},\begin{bmatrix}
\Psi^{L}(1)\\
\Psi^{R}(1)\end{bmatrix},\cdots\right] \in(\mathbb{C}^{2})^{\mathbb{Z}},
\end{align*}
then 
\begin{align*}
\phi(\Psi) = {}^T\! 
\left[\ldots, 
|\Psi^{L}(-1)|^2 + |\Psi^{R}(-1)|^2, 
|\Psi^{L}(0)|^2 + |\Psi^{R}(0)|^2, 
|\Psi^{L}(1)|^2 + |\Psi^{R}(1)|^2, \ldots \right] 
\in \mathbb{R}_{+}^{\mathbb{Z}}.
\end{align*}
That is, for any $x \in \ZM$, 
\begin{align*}
\phi(\Psi) (x) = |\Psi^{L}(x)|^2 + |\Psi^{R}(x)|^2.
\end{align*}
Sometime we identify $\phi(\Psi(x))$ with $\phi(\Psi) (x)$. Moreover we define the measure of the QW at position $x$ by
\begin{align*}
\mu(x)=\phi(\Psi(x)) \quad (x \in \ZM).
\end{align*}
Now we are ready to introduce the set of stationary measure:  
\begin{align*}
&{\cal M}_{s} = {\cal M}_s (U)
\\
&= \left\{ \phi(\Psi_{0})\in\mathbb{R}_{+}^{\mathbb{Z}} \setminus \{ \boldsymbol{0} \} : there\;exists\;\Psi_{0}\;such\;that\;\;\phi((U^{(s)})^{n}\Psi_{0})=\phi(\Psi_{0})\;for\;any\;n\geq 0 \right\},
\end{align*}
where $\boldsymbol{0}$ is the zero vector. We call the element of ${\cal M}_{s}$ the stationary measure of the QW.

Next we consider the eigenvalue problem of the QW:
\begin{align}
U^{(s)} \Psi = \lambda \Psi \quad (\lambda \in \mathbb{C}).
\label{samui}
\end{align}
Remark that $|\lambda|=1$, since $U^{(s)}$ is unitary. We sometime write $\Psi=\Psi^{(\lambda)}$ in order to emphasize the dependence on eigenvalue $\lambda$. Then we see that $\phi (\Psi^{(\lambda)}) \in {\cal M}_s$. Moreover we introduce 
\begin{align*} 
{\cal W}^{(\lambda)} 
= \left\{ \Psi^{(\lambda)} \in \CM^{\ZM} \setminus \{ \boldsymbol{0} \} : U^{(s)} \Psi^{(\lambda)} = \lambda \Psi^{(\lambda)} \right\}.
\end{align*}
We see that Eq.(\ref{samui}) is equivalent to 
\begin{align}
\lambda \Psi^{L}(x) 
&= a \Psi^{L}(x+1) + b \Psi^{R} (x+1),
\label{yokoyama}
\\
\lambda \Psi^{R}(x)
&= c \Psi^{L}(x-1) + d \Psi^{R}(x-1),
\label{taikan} 
\end{align}
for any $x \in \ZM$.

Let $\mu_n (x)$ be the measure of the QW at position $x$ and at time $n$, i.e., \begin{align*}
\mu_n (x)=\phi(\Psi_n(x)) \quad (x \in \ZM).
\end{align*}
If $\lim_{n \to \infty} \mu_n (x)$ exists for any $x \in \ZM$, then we define the limit measure $\mu_{\infty} (x)$ by
\begin{align*}
\mu_{\infty} (x) = \lim_{n \to \infty} \mu_n (x) \quad (x \in \ZM).
\end{align*}
Moreover we put the time-averaged limit measure 
\begin{align*}
\overline{\mu}_{\infty} (x) = \lim_{T \to \infty} \frac{1}{T} \sum_{n=0}^{T-1} \mu_n (x) \quad (x \in \ZM),
\end{align*}
if the right-hand side of the above equation exists.

Let $\Psi_0 ^{\{0\}}$ be the initial state for the QW starting from the origin; \begin{align*}
\Psi_0 ^{\{0\}} 
&= \Psi_0 ^{\{0\}} (\alpha, \beta) = {}^T \left[ \ldots, 
\begin{bmatrix} \Psi^{L} (-2) \\ \Psi^{R} (-2) \end{bmatrix}, 
\begin{bmatrix} \Psi^{L} (-1) \\ \Psi^{R} (-1) \end{bmatrix}, 
\begin{bmatrix} \Psi^{L} (0)  \\ \Psi^{R} (0)  \end{bmatrix}, 
\begin{bmatrix} \Psi^{L} (1)  \\ \Psi^{R} (1)  \end{bmatrix}, 
\begin{bmatrix} \Psi^{L} (2)  \\ \Psi^{R} (2)  \end{bmatrix}, 
\ldots \right],
\\
&= {}^T \left[ \ldots, 
\begin{bmatrix} 0 \\ 0 \end{bmatrix}, 
\begin{bmatrix} 0 \\ 0 \end{bmatrix}, 
\begin{bmatrix} \alpha  \\ \beta  \end{bmatrix}, 
\begin{bmatrix} 0 \\ 0  \end{bmatrix}, 
\begin{bmatrix} 0 \\ 0  \end{bmatrix}, 
\ldots \right],
\end{align*}
with $|\alpha|^2+|\beta|^2=1$. Then we introduce 
\begin{align*} 
{\cal M}_{\infty} ^{\{0\}}
&= 
\left\{ \mu_{\infty} = \mu_{\infty} ^{\Psi_0 ^{\{0\}}} \in \RM_+^{\ZM} \setminus \{ \boldsymbol{0} \} : \Psi_0 ^{\{0\}} = \Psi_0 ^{\{0\}} (\alpha, \beta) \in \CM^{\ZM} \> \hbox{with} \> |\alpha|^2+|\beta|^2=1 \right\},
\\
\overline{{\cal M}}_{\infty} ^{\{0\}}
&= 
\left\{ \overline{\mu}_{\infty} = \overline{\mu}_{\infty} ^{\Psi_0 ^{\{0\}}} \in \RM_+^{\ZM} \setminus \{ \boldsymbol{0} \} : \Psi_0 ^{\{0\}} = \Psi_0 ^{\{0\}} (\alpha, \beta) \in \CM^{\ZM} \> \hbox{with} \> |\alpha|^2+|\beta|^2=1 \right\},
\\
{\cal M}^{(w,\{0\})}
&= 
\left\{ w \hbox{-} \lim_{n \to \infty} \frac{X_n}{n} : \Psi_0 ^{\{0\}} = \Psi_0 ^{\{0\}} (\alpha, \beta) \in \CM^{\ZM} \> \hbox{with} \> |\alpha|^2+|\beta|^2=1 \right\},
\end{align*}
where $w$-$\lim_{n \to \infty} X_n/n$ stands for the weak convergence limit measure for $X_n/n$ if it exists.

From now on we will show that $\mu_{u,\ZM}^{(c)} \in {\cal M}_s (U) \> (c >0)$ for $a=0$ (Sect. \ref{azero}), $b=0$ (Sect. \ref{bzero}), and $abcd \not=0$ (Sect. \ref{abcdzero}).

\section{Case $a=0$ \label{azero}}
In this case, $U$ can be expressed as 
\begin{align*}
U=
\left[
\begin{array}{cc}
0 & e^{i \eta}  \\
- \triangle e^{-i \eta} & 0
\end{array}
\right],
\end{align*}
where $\eta \in [0, 2 \pi)$ and $\triangle (= \det U) \in \mathbb{C}$ with $|\triangle|=1$. 

From Eqs.(\ref{yokoyama}) and (\ref{taikan}), we get
\begin{align*}
\lambda \Psi^{L}(x) 
&=  e^{i \eta}  \Psi^{R} (x+1),
%\label{yokoyama1}
\\
\lambda \Psi^{R}(x)
&= - \triangle e^{-i \eta} \Psi^{L}(x-1).
%\label{taikan1} 
\end{align*}
By these equations, we see that for any $x \in \ZM$,
\begin{align*}
\left( 1 + \frac{\triangle}{\lambda^2} \right) \Psi^{j}(x) = 0 \quad (j=L,R).
\end{align*}
From this, we get $\triangle = - \lambda^2$. Let $\lambda_{\pm} = \pm i \sqrt{\triangle},$ where the sign is chosen in a suitable way. As an initial state, we consider $\Psi_0^{(\pm)}$ corresponding to $\lambda_{\pm}$ as follows;
\begin{align}
\Psi_0^{(\pm)} = {}^T \left[ \ldots, 
\begin{bmatrix} \Psi^{(\pm,L)} (-2) \\ \Psi^{(\pm,R)} (-2) \end{bmatrix}, 
\begin{bmatrix} \Psi^{(\pm,L)} (-1) \\ \Psi^{(\pm,R)} (-1) \end{bmatrix}, 
\begin{bmatrix} \Psi^{(\pm,L)} (0)  \\ \Psi^{(\pm,R)} (0)  \end{bmatrix}, 
\begin{bmatrix} \Psi^{(\pm,L)} (1)  \\ \Psi^{(\pm,R)} (1)  \end{bmatrix}, 
\begin{bmatrix} \Psi^{(\pm,L)} (2)  \\ \Psi^{(\pm,R)} (2)  \end{bmatrix}, 
\ldots \right].
\label{huyumiHayaku}
\end{align}
Here for any $x \in \ZM$, 
\begin{align}
\Psi^{(\pm,L)} (2x) &= \alpha, \>\> \Psi^{(\pm,R)} (2x) = \beta, 
\nonumber
\\
\Psi^{(\pm,L)} (2x-1) &= \frac{e^{i \eta}}{\lambda_{\pm}} \beta, \>\> \Psi^{(\pm,R)} (2x+1) = - \frac{\triangle e^{-i \eta}}{\lambda_{\pm}} \alpha = \lambda_{\pm} e^{-i \eta} \alpha,
\label{huyumiHayaku2}
\end{align}
where $\alpha, \> \beta \in \CM$ with $\alpha \beta \not= 0$. In fact, we have 
\begin{align*}
U^{(s)} \Psi_0^{(\pm)} = \lambda_{\pm} \Psi_0^{(\pm)}.
\end{align*}
Then $\Psi_0^{(\pm)} \in {\cal W}^{(\lambda_{\pm})}$. Therefore
\begin{align}
(U^{(s)})^n \Psi_0^{(\pm)} = \lambda_{\pm}^n \Psi_0^{(\pm)}.
\label{huyumiHayaku3}
\end{align}

For a special case with $\eta =0, \> \triangle=-1$, i.e., 
\begin{align*}
U=
\left[
\begin{array}{cc}
0 & 1  \\
1 & 0
\end{array}
\right],
\end{align*}
$\lambda=1,$ and $\alpha = \beta = \psi (\not=0)$, we have
\begin{align*}
U^{(s)} \Psi = \Psi,
\end{align*}
where $\Psi$ is the uniform amplitude with parameter $\psi$, that is, $\Psi = \Psi_{u, \ZM}^{(\psi)}$. In fact, this QW is the two-state Grower walk considered in Sect. \ref{gw}.

Let $\mu_n ^{(\Psi_0^{(\pm)})} = \phi ((U^{(s)})^n \Psi_0^{(\pm)})$ and  
\begin{align*}
\mu_n ^{(\Psi_0^{(\pm)})}
= {}^T \left[ \ldots, \mu_n^{(\Psi_0^{(\pm)})} (-2), \mu_n^{(\Psi_0^{(\pm)})} (-1), \mu_n^{(\Psi_0^{(\pm)})} (0), \mu_n^{(\Psi_0^{(\pm)})} (1), \mu_n^{(\Psi_0^{(\pm)})} (2), \ldots \right].
\end{align*}
From Eqs.(\ref{huyumiHayaku}), (\ref{huyumiHayaku2}), and (\ref{huyumiHayaku3}), we obtain  
\begin{align*}
\mu_n ^{(\Psi_0^{(\pm)})} = {}^T \left[ \ldots, |\alpha|^2+|\beta|^2, |\alpha|^2+|\beta|^2, |\alpha|^2+|\beta|^2, |\alpha|^2+|\beta|^2, |\alpha|^2+|\beta|^2, \ldots \right].
\end{align*}
Therefore we see that for any $n \ge 0$, $\mu_n ^{(\Psi_0^{(\pm)})} = \mu_0 ^{(\Psi_0^{(\pm)})}$. So $\mu_0 ^{(\Psi_0^{(\pm)})}$ becomes the stationary measure, that is, $\mu_0 ^{(\Psi_0^{(\pm)})} \in {\cal M}_s (U)$. Moreover $\mu_n ^{(\Psi_0^{(\pm)})} (x) = |\alpha|^2+|\beta|^2 \> (x \in \ZM)$. So $\mu_0^{(\Psi_0^{(\pm)})}$ is the uniform measure, i.e., $\mu_0^{(\Psi_0^{(\pm)})} = \mu_{u, \ZM}^{(c)}$ with $c=|\alpha|^2+|\beta|^2$. Then we have $\mu_0^{(\Psi_0^{(\pm)})} = \mu_{u, \ZM}^{(c)} \in {\cal M}_s (U).$

On the other hand, we consider the QW starting from the origin: 
\begin{align*}
\Psi_0 ^{\{0\}} 
&= {}^T \left[ \ldots, 
\begin{bmatrix} \Psi^{L} (-2) \\ \Psi^{R} (-2) \end{bmatrix}, 
\begin{bmatrix} \Psi^{L} (-1) \\ \Psi^{R} (-1) \end{bmatrix}, 
\begin{bmatrix} \Psi^{L} (0)  \\ \Psi^{R} (0)  \end{bmatrix}, 
\begin{bmatrix} \Psi^{L} (1)  \\ \Psi^{R} (1)  \end{bmatrix}, 
\begin{bmatrix} \Psi^{L} (2)  \\ \Psi^{R} (2)  \end{bmatrix}, 
\ldots \right],
\\
&= {}^T \left[ \ldots, 
\begin{bmatrix} 0 \\ 0 \end{bmatrix}, 
\begin{bmatrix} 0 \\ 0 \end{bmatrix}, 
\begin{bmatrix} \alpha  \\ \beta  \end{bmatrix}, 
\begin{bmatrix} 0 \\ 0  \end{bmatrix}, 
\begin{bmatrix} 0 \\ 0  \end{bmatrix}, 
\ldots \right],
\end{align*}
with $|\alpha|^2+|\beta|^2=1$. Then the definition of the QW implies that for any $n \ge 0$, 
\begin{align*}
\mu_{2n} = \delta_0, \quad \mu_{2n+1} = |\beta|^2 \delta_{-1} + |\alpha|^2 \delta_1.
\end{align*}
Here $\delta_x$ denotes the delta measure at position $x \in \ZM$. So we should remark that for fixed $x \in \{-1,0,1\}$, $\mu_n (x)$ does not converge as $n \to \infty$. However $\lim_{n \to \infty} \mu_n (x) = 0$ for fixed $x \in \ZM \setminus \{-1,0,1\}$. Moreover we have 
\begin{align*}
\lim_{n \to \infty} E(e^{i \xi X_n/n}) = 1, 
\end{align*}
where $\xi \in \RM$, since we see that if $n$ is even, then $E(e^{i \xi X_n/n})=1$, if $n$ is odd, then 
\begin{align*}
E(e^{i \xi X_n/n}) 
= \cos ( \xi/n ) + i (|\beta|^2 - |\alpha|^2) \sin (\xi/n).
\end{align*}
Thus
\begin{align*}
X_n /n \quad \Rightarrow \quad \delta_0,
\end{align*}
where $\Rightarrow$ means the weak convergence. Therefore we have
\begin{align*}
{\cal M}_{\infty} ^{\{0\}} 
&= \emptyset, 
\quad 
\overline{{\cal M}}_{\infty} ^{\{0\}}
= \left\{  \frac{1}{2} \left( |\beta|^2 \delta_{-1} + \delta_{0} + |\alpha|^2 \delta_1 \right) : \alpha, \beta \in \CM \> \hbox{with} \> |\alpha|^2+|\beta|^2=1 \right\},
\\
{\cal M}^{(w,\{0\})}
&= \left\{ \delta_0 \right\}.
\end{align*}
Thus we see that for any $c>0$ and $-1 \le c_1<c_2 \le 1$, 
\begin{align*}
\mu_{u,\ZM} ^{(c)} \not\in \overline{{\cal M}}_{\infty} ^{\{0\}}, \quad \mu_{u,(c_1,c_2)} ^{(c)} \not\in {\cal M}^{(w,\{0\})}.
\end{align*}

\section{Case $b=0$ \label{bzero}}
In this case, we see that $U$ can be written as 
\[
U=
\left[
\begin{array}{cc}
e^{i \eta} & 0 \\
0 & \triangle e^{-i \eta} 
\end{array}
\right],
\]
where $\eta \in [0, 2 \pi)$ and $\triangle (= \det U) \in \mathbb C$ with $|\triangle|=1$. From Eqs.(\ref{yokoyama}) and (\ref{taikan}), we get
\begin{align*}
\lambda \Psi^{L}(x) 
&=  e^{i \eta}  \Psi^{L} (x+1),
%\label{yokoyama2}
\\
\lambda \Psi^{R}(x)
&= \triangle e^{-i \eta} \Psi^{R}(x-1).
%\label{taikan2} 
\end{align*}
By these equations, we see that for any $x \in \ZM$,
\begin{align*}
\Psi^{L}(x) =  (\lambda e^{-i \eta})^x  \> \alpha, 
\quad 
\Psi^{R}(x) =  (\overline{\lambda} \triangle e^{-i \eta})^x \> \beta,
\end{align*}
where $\alpha, \> \beta \in \CM$ with $\alpha \beta \not= 0$. If we take the above $\Psi$ as the initial state $\Psi_0$, then we have 
\begin{align*}
\Psi_n = (U^{(s)})^n \Psi_0 = \lambda^n \Psi_0.
\end{align*}

For a special case with $\eta =0, \> \triangle=1$, i.e., 
\begin{align*}
U=
\left[
\begin{array}{cc}
1 & 0  \\
0 & 1
\end{array}
\right],
\end{align*}
$\lambda=1,$ and $\alpha = \beta = \psi (\not=0)$, we have
\begin{align*}
U^{(s)} \Psi = \Psi,
\end{align*}
where $\Psi$ is the uniform amplitude with parameter $\psi$, that is, $\Psi = \Psi_{u, \ZM}^{(\psi)}$.

Then we have the measure $\mu_n$ at time $n$ as follows:
\begin{align*}
\mu_n (x)
&= |\Psi_n^{L}(x)|^2 + |\Psi_n^{R}(x)|^2 = |\lambda|^{2n} \left( |\Psi_0^{L}(x)|^2 + |\Psi_0^{R}(x)|^2 \right) = |\alpha|^2 + |\beta|^2.
%\label{taiyouda}
\end{align*}
So $\mu_0$ becomes the stationary measure, that is, $\mu_0 \in {\cal M}_s (U)$. Moreover $\mu_0 (x) = |\alpha|^2 + |\beta|^2 \> (x \in \ZM)$. So $\mu_0$ is the uniform measure, i.e., $\mu_0 = \mu_{u, \ZM}^{(c)}$ with $c=|\alpha|^2 + |\beta|^2$. Then we have $\mu_0 = \mu_{u, \ZM}^{(c)} \in {\cal M}_s (U).$

As in the case of $a=0$, we consider the following initial state,  
\begin{align*}
\Psi_0 ^{\{0\}} 
&= {}^T \left[ \ldots, 
\begin{bmatrix} \Psi^{L} (-2) \\ \Psi^{R} (-2) \end{bmatrix}, 
\begin{bmatrix} \Psi^{L} (-1) \\ \Psi^{R} (-1) \end{bmatrix}, 
\begin{bmatrix} \Psi^{L} (0)  \\ \Psi^{R} (0)  \end{bmatrix}, 
\begin{bmatrix} \Psi^{L} (1)  \\ \Psi^{R} (1)  \end{bmatrix}, 
\begin{bmatrix} \Psi^{L} (2)  \\ \Psi^{R} (2)  \end{bmatrix}, 
\ldots \right],
\\
&= {}^T \left[ \ldots, 
\begin{bmatrix} 0 \\ 0 \end{bmatrix}, 
\begin{bmatrix} 0 \\ 0 \end{bmatrix}, 
\begin{bmatrix} \alpha  \\ \beta  \end{bmatrix}, 
\begin{bmatrix} 0 \\ 0  \end{bmatrix}, 
\begin{bmatrix} 0 \\ 0  \end{bmatrix}, 
\ldots \right],
\end{align*}
with $|\alpha|^2+|\beta|^2=1$. Then the definition of the QW implies 
that 
\begin{align}
\mu_{n} = |\alpha|^2 \delta_{-n} + |\beta|^2 \delta_n \quad (n \ge 0).
\label{hukushima}
\end{align}
So we get
\begin{align*}
\lim_{n \to \infty} \mu_{n} = \boldsymbol{0},
\end{align*}
that is, $\lim_{n \to \infty} \mu_{n} (x) = 0$ for any $x \in \ZM$. Equation (\ref{hukushima}) implies 
\begin{align*}
X_n /n \quad \Rightarrow \quad |\alpha|^2 \delta_{-1} + |\beta|^2 \delta_{1}.
\end{align*}
Therefore we have
\begin{align*}
{\cal M}_{\infty} ^{\{0\}} 
&= \overline{{\cal M}}_{\infty} ^{\{0\}}
= \emptyset, 
\\ 
{\cal M}^{(w,\{0\})}
&= \left\{ |\alpha|^2 \delta_{-1} + |\beta|^2 \delta_1 : \alpha, \beta \in \CM \> \hbox{with} \> |\alpha|^2+|\beta|^2=1 \right\}.
\end{align*}
Thus we see that for any $c>0$ and $-1 \le c_1<c_2 \le 1$, 
\begin{align*}
\mu_{u,(c_1,c_2)} ^{(c)} \not\in {\cal M}^{(w,\{0\})}.
\end{align*}

\section{Case $abcd \not=0$ \label{abcdzero}}
Let $\Psi(x)={}^T \> [\Psi^{L}(x),\> \Psi^{R}(x)] \>\> (x \in \mathbb{Z})$
be the amplitude of the model at position $x$. Here we introduce the generating functions for $\Psi^{L}(x)$ and $\Psi^{R}(x)$:
\begin{align*}
f^{j}(z)=\sum_{x \in  \mathbb{Z}} \Psi^{j}(x)z^{x} \quad(j=L,R).
\end{align*}
From Eqs.(\ref{yokoyama}) and (\ref{taikan}), we obtain the following lemma.
\begin{lem}
We have
\begin{align*}
A f (z) = 
\begin{bmatrix}
0 \\
0
\end{bmatrix},
\end{align*}
where 
\begin{align*}
A = 
\begin{bmatrix}
\lambda - \dfrac{a}{z} & -\dfrac{b}{z} \\
- cz & \lambda - dz
\end{bmatrix}, 
\qquad 
f (z) = 
\begin{bmatrix}
f^{L} (z) \\ 
f^{R} (z)
\end{bmatrix}.
\end{align*}
\end{lem}
Then we have
\begin{align*}
\det A = - \frac{d \lambda}{z} h(z),
\end{align*}
where
\begin{align*}
h(z) = z^{2}- \frac{1}{d} \left(\lambda + \frac{\triangle}{\lambda} \right) z + \frac{a}{d}.
\end{align*}
Let $\phi \in (0, \pi/2)$ satisfy $\cos \phi = |a|, \> \sin \phi = \sqrt{1-|a|^2}$. Put $\xi \in [0, 2 \pi)$ with $\triangle = e^{i \xi}$. For the following four $\lambda$'s, $h(z)$ has double roots.
\begin{align*}
\lambda_1 = e^{i(\phi+(\xi/2))}, \> \lambda_2 = e^{i(-\phi+(\xi/2))}, \> \lambda_3 =  - \lambda_1, \> \lambda_4 = - \lambda_2.
\end{align*}

Moreover Eqs.(\ref{yokoyama}) and (\ref{taikan}) imply that $\Psi^{L}(x)$ and $\Psi^{R}(x)$ satisfy the following same equation:
\begin{align}
a_{x+2} - \frac{1}{a} \left(\lambda + \frac{\triangle}{\lambda} \right) a_{x+1} + \frac{d}{a} a_x = 0,
\label{aida}
\end{align}
for any $x \in \mathbb{Z}$. We put 
\begin{align}
\Psi^{L}(x) = A \gamma^x \quad  (x \in \mathbb{Z}),
\label{hayami}
\end{align}
where $A (\not=0) \in \mathbb{C}$. Here $\gamma \in \mathbb{C}$ is the double roots of the following characteristic polynomial for difference equation (\ref{aida}):
\begin{align*}
x^2 - \frac{1}{a} \left(\lambda + \frac{\triangle}{\lambda} \right) x + \frac{d}{a} = 0,
\end{align*}
Then we have
\begin{align}
\gamma = \frac{\lambda + \triangle \overline{\lambda}}{2a}.
\label{makoto}
\end{align}
From Eqs.(\ref{yokoyama}) and (\ref{hayami}), we have
\begin{align*}
\Psi^{R}(x) = \frac{A}{b} \left( \frac{\lambda - \triangle \overline{\lambda}}{2} \right) \gamma^{x-1} \quad  (x \in \mathbb{Z}).
%\label{gyosyu}
\end{align*}
Then we see that for any $x \in \mathbb{Z}$,
\begin{align*}
\Psi^{L}(x) = A \gamma^x, \quad \Psi^{R}(x) = \frac{A}{b} \left( \frac{\lambda - \triangle \overline{\lambda}}{2} \right) \gamma^{x-1}. 
\end{align*}
In fact, $\Psi^{L}(x)$ and $\Psi^{R}(x)$ satisfy Eq.(\ref{taikan}). Therefore we obtain the following result.

\begin{pro}
For the QW with $abcd \not=0$, we see that
\begin{align*}
\Psi(x) = 
\begin{bmatrix}
\Psi^{L}(x) \\
\Psi^{R}(x)
\end{bmatrix}
=
\begin{bmatrix}           
A \gamma^x, \\
\dfrac{A}{b} \left( \dfrac{\lambda - \triangle \overline{\lambda}}{2} \right) \gamma^{x-1}
\end{bmatrix}
\quad  (x \in \mathbb{Z})
\end{align*}
satisfies
\begin{align*}
U^{(s)} \Psi = \lambda \Psi.
\end{align*}
\end{pro}

If we take the above $\Psi$ as the initial state $\Psi_0$, then we have 
\begin{align*}
\Psi_n = (U^{(s)})^n \Psi_0 = \lambda^n \Psi_0.
\end{align*}
Therefore we have the measure $\mu_n$ at time $n$ as follows:
\begin{align}
\mu_n (x) 
&= |\Psi_n^{L}(x)|^2 + |\Psi_n^{R}(x)|^2 = |\lambda|^{2n} \left( |\Psi_0^{L}(x)|^2 + |\Psi_0^{R}(x)|^2 \right) 
\nonumber
\\
&= |A|^2 \left( |\gamma|^2 + \frac{|\lambda - \triangle \overline{\lambda}|^2}{4|b|^2} \right) |\gamma|^{2(x-1)}.
\label{taiyou}
\end{align}
From now on we compute $|\gamma|$. Eq.(\ref{makoto}) gives 
\begin{align}
|\gamma|^2 = \frac{ 1 + \Re (\triangle \overline{\lambda}^2)}{2 |a|^2},
\label{akiba}
\end{align}
where $\Re (z)$ is the real part of $z \in \CM$. On the other hand, for any $\lambda_k \> (k=1,2,3,4)$, we see that
\begin{align}
\Re (\triangle \overline{\lambda}^2) = \cos (2 \phi) = 2 |a|^2 -1.
\label{otaku}
\end{align}
Combining Eq.(\ref{akiba}) with Eq.(\ref{otaku}) implies $|\gamma|=1$.

Moreover, in a similar way, 
\begin{align}
\frac{|\lambda - \triangle \overline{\lambda}|^2}{4|b|^2} 
= \frac{1 - \Re (\triangle \overline{\lambda}^2)}{2 |b|^2}
=\frac{2(1-|a|^2))}{2|b|^2} = 1.
\label{kozou}
\end{align}
From Eq.(\ref{taiyou}), $|\gamma|=1$, and Eq.(\ref{kozou}), we obtain 
\begin{align*}
\mu_n (x) = 2 |A|^2,
\end{align*}
for any $n \ge 0$ and $x \in \mathbb{Z}$. So let $\mu=\mu_n$. Then this $\mu$ becomes stationary and uniform measure of the QW defined by $U$ with $abcd \not=0$. Therefore $\mu = \mu^{(c)}_{u,\ZM} \in {\cal M}_s (U)$ with $c= 2 |A|^2$.

Here we consider the case of the QW determined by $U=U(\theta)$ with $0 < \theta < \pi/2$. In this case, we have $\phi = \theta$. Moreover $\xi = \pi$, since $\triangle = \det U(\theta) = -1$. Let 
\begin{align*}
\gamma_k = \frac{\lambda_k + \triangle \overline{\lambda_k}}{2 \cos \theta} \quad (k=1,2,3,4).
\end{align*}
So we have
\begin{align*}
\gamma_1 = \gamma_2 = i, \quad \gamma_3 = \gamma_4 = -i.
\end{align*}
For $k=1,2,3,4$, we put
\begin{align*}
\Psi^{(k)} (x) = 
\begin{bmatrix}
\Psi^{(k,L)}(x) \\
\Psi^{(k,R)}(x)
\end{bmatrix}.
\end{align*}
Therefore
\begin{align*}
\Psi^{(1)} (x) 
&= 
\begin{bmatrix}
A i^x \\
- A i^{x-1}
\end{bmatrix},
\quad 
\Psi^{(2)} (x) = 
\begin{bmatrix}
A i^x \\
A i^{x-1}
\end{bmatrix},
\nonumber
\\
\Psi^{(3)} (x) 
&= 
\begin{bmatrix}
A (-i)^x \\
A (-i)^{x-1}
\end{bmatrix},
\quad 
\Psi^{(4)} (x) = 
\begin{bmatrix}
A (-i)^x \\
-A (-i)^{x-1}
\end{bmatrix}.
\end{align*}
Then $\Psi^{(k)} \in {\cal W}^{(\lambda_k)}.$ Thus $\mu_n (x) = 2 |A|^2$ for any $n \ge 0$ and $x \in \ZM$.

On the other hand, for the following initial state, 
\begin{align*}
\Psi_0 ^{\{0\}} 
&= {}^T \left[ \ldots, 
\begin{bmatrix} \Psi^{L} (-2) \\ \Psi^{R} (-2) \end{bmatrix}, 
\begin{bmatrix} \Psi^{L} (-1) \\ \Psi^{R} (-1) \end{bmatrix}, 
\begin{bmatrix} \Psi^{L} (0)  \\ \Psi^{R} (0)  \end{bmatrix}, 
\begin{bmatrix} \Psi^{L} (1)  \\ \Psi^{R} (1)  \end{bmatrix}, 
\begin{bmatrix} \Psi^{L} (2)  \\ \Psi^{R} (2)  \end{bmatrix}, 
\ldots \right],
\\
&= {}^T \left[ \ldots, 
\begin{bmatrix} 0 \\ 0 \end{bmatrix}, 
\begin{bmatrix} 0 \\ 0 \end{bmatrix}, 
\begin{bmatrix} \alpha  \\ \beta  \end{bmatrix}, 
\begin{bmatrix} 0 \\ 0  \end{bmatrix}, 
\begin{bmatrix} 0 \\ 0  \end{bmatrix}, 
\ldots \right],
\end{align*}
with $|\alpha|^2+|\beta|^2=1$, Konno \cite{Konno2002, Konno2005} proved
\begin{align*}
{X_n \over n} \quad \Rightarrow \quad Z \qquad (n \to \infty),
\end{align*}
where $Z$ has the following density function:
\begin{align*}
f(x) = f(x; {}^T[\alpha, \beta]) = \left\{ 1 - C(a,b; \alpha, \beta)x \right\} f_K(x;|a|).
\end{align*}
Here
\begin{align*}
C(a,b; \alpha, \beta) 
&= |\alpha|^2 - |\beta|^2 + {a \alpha \overline{b \beta} + \overline{a \alpha} b \beta \over |a|^2 },
\\
f_K(x;|a|) 
&= { \sqrt{1 - r^2} \over \pi (1 - x^2) \sqrt{r^2 - x^2}} \> I_{(-r,r)}(x),
\end{align*}
where $I_A (x) =1 \> (x \in A), \> =0 \> (x \not\in A)$. That is,
\begin{align*}
\lim_{n \to \infty} P\left( u \le \frac{X_n}{n} \le v \right)= \int_u^v f(x)dx.
\end{align*}
As a corollary, one obtains 
\begin{align*}
\lim_{n \to \infty} \mu_n (x) = 0,
\end{align*}
for any fixed $x \in Z$. Therefore we see that
\begin{align*}
{\cal M}_{\infty} ^{\{0\}} 
&= \overline{{\cal M}}_{\infty} ^{\{0\}}
= \emptyset, 
\\ 
{\cal M}^{(w,\{0\})}
&= \left\{ f(x; {}^T[\alpha, \beta])dx : \alpha, \beta \in \CM \> \hbox{with} \> |\alpha|^2+|\beta|^2=1 \right\}.
\end{align*}
Thus we see that for any $c>0$ and $-1 \le c_1<c_2 \le 1$, 
\begin{align*}
\mu_{u,(c_1,c_2)} ^{(c)} \not\in {\cal M}^{(w,\{0\})}.
\end{align*}

\section{Three-State Grover Walk \label{gw}}
As in a similar argument for the two-state QW, we consider the stationary measure of the three-state Grover walk determined by the unitary matrix $U$:
\begin{align*}
U=
\frac{1}{3}
\left[
\begin{array}{ccc}
-1 & 2  &   2 \\
2  & -1 &   2 \\
2  & 2  &  -1 \\
\end{array}
\right].
\end{align*}
To define the dynamics of the walk, we divide $U$ into three matrices:
\begin{align*}
U_{L}=
\frac{1}{3}
\left[
\begin{array}{ccc}
-1 & 2 &  2 \\
0 & 0 &  0 \\
0 & 0 &  0 \\
\end{array}
\right], \quad 
U_{0}=
\frac{1}{3}
\left[
\begin{array}{ccc}
0 & 0 &  0 \\
2 & -1 & 2 \\
0 & 0 &  0 \\
\end{array}
\right], \quad
U_{R}=
\frac{1}{3}
\left[
\begin{array}{ccc}
0 & 0 &  0 \\
0 & 0 &  0 \\
2 & 2 &  -1 \\
\end{array}
\right],
%\label{konno-eqn:U3}
\end{align*}
with $U=U_{L}+U_{0}+U_{R}$. The important point is that $U_L$ (resp. $U_R$) represents that the walker moves to the left (resp. right) at any position at each time step. $U_0$ represents that the walker stays at the same position.

Let $\Psi_{n}$ denote the amplitude at time $n$ of the Grover walk:
\begin{align*}
\Psi_{n}= {}^T\![\cdots,\Psi_{n}^{L}(-1),\Psi_{n}^{0}(-1),\Psi_{n}^{R}(-1),\Psi_{n}^{L}(0),\Psi_{n}^{0}(0),\Psi_{n}^{R}(0),\Psi_{n}^{L}(1),\Psi_{n}^{0}(1),\Psi_{n}^{R}(1),\cdots ].
\end{align*}
As in the two-state case, the evolution is given by 
\begin{align*}
\Psi_{n+1} (x) = U_{L} \Psi_{n} (x+1) + U_{0} \Psi_{n} (x) + U_{R} \Psi_{n} (x-1).
\end{align*}
Put
\begin{align*}
U^{(s)}=\begin{bmatrix}
\ddots&\vdots&\vdots&\vdots&\vdots&\cdots\\
\cdots&U_0&U_L&O&O&O\cdots\\
\cdots&U_R&U_0&U_L&O&O\cdots\\
\cdots&O&U_R&U_0&U_L&O\cdots\\
\cdots&O&O&U_R&U_0&U_L\cdots\\
\cdots&O&O&O&U_R&U_0\cdots\\
\cdots&\vdots&\vdots&\vdots&\vdots&\ddots
\end{bmatrix}\;\;\;
with\;\;\;O=\begin{bmatrix}0&0&0\\0&0&0\\0&0&0\end{bmatrix}.
\end{align*}
Then the state of the QW at time $n$ is determined by $\Psi_{n}=(U^{(s)})^{n}\Psi_{0}$ for any $n\geq0$. Here we introduce a map $\phi:(\mathbb{C}^{3})^{\mathbb{Z}}\rightarrow \mathbb{R}_{+}^{\mathbb{Z}}$ such that if
\begin{align*}
\Psi= {}^T\!\left[\cdots,\begin{bmatrix}
\Psi^{L}(-1)\\
\Psi^{0}(-1)\\
\Psi^{R}(-1)\end{bmatrix},\begin{bmatrix}
\Psi^{L}(0)\\
\Psi^{0}(0)\\
\Psi^{R}(0)\end{bmatrix},\begin{bmatrix}
\Psi^{L}(1)\\
\Psi^{0}(1)\\
\Psi^{R}(1)\end{bmatrix},\cdots\right]\in(\mathbb{C}^{3})^{\mathbb{Z}},
\end{align*}
then 
\begin{align*}
\phi(\Psi) 
&= {}^T\! 
\left[ \ldots, 
|\Psi^{L}(-1)|^2 + |\Psi^{0}(-1)|^2 + |\Psi^{R}(-1)|^2, 
|\Psi^{L}(0)|^2 + |\Psi^{0}(0)|^2 + |\Psi^{R}(0)|^2, 
\right.
\\
& \qquad \qquad \left. |\Psi^{L}(1)|^2 + |\Psi^{0}(1)|^2 + |\Psi^{R}(1)|^2, \ldots 
\right] \in \mathbb{R}_{+}^{\mathbb{Z}}.
\end{align*}
That is, for any $x \in \ZM$, 
\begin{align*}
\phi(\Psi) (x) = \phi(\Psi(x))= |\Psi^{L}(x)|^2 + |\Psi^{0}(x)|^2 + |\Psi^{R}(x)|^2.
\end{align*}
Moreover we define the measure of the QW at position $x$ by
\begin{align*}
\mu(x)=\phi(\Psi(x)) \quad (x \in \ZM).
\end{align*}
As in the case of two-state QW, we consider the stationary measure. First we consider the eigenvalue problem:
\begin{align}
U^{(s)} \Psi = \lambda \Psi,
\label{mishima}
\end{align}
where $\lambda\in\mathbb{C}$ with $|\lambda|=1$. Then Eq.(\ref{mishima}) is equivalent to
\begin{align}
\lambda \Psi^{L}(x) 
&= -\frac{1}{3} \Psi^{L}(x+1) + \frac{2}{3} \Psi^{0}(x+1) + \frac{2}{3} \Psi^{R} (x+1),
\label{yokoyama3}
\\
\lambda \Psi^{0}(x) 
&= \frac{2}{3} \Psi^{L}(x) - \frac{1}{3} \Psi^{0}(x) +\frac{2}{3} \Psi^{R} (x),
\label{yokoyamama3}
\\
\lambda \Psi^{R}(x)
&= \frac{2}{3} \Psi^{L}(x-1) +\frac{2}{3} \Psi^{0}(x-1) -\frac{1}{3} \Psi^{R} (x-1),
\label{taikan3} 
\end{align}
for any $x \in \ZM$. Here we introduce the generating functions for $\Psi^{L}(x), \> \Psi^{0}(x)$ and $\Psi^{R}(x)$:
\begin{align*}
f_G ^{j}(z)=\sum_{x \in  \mathbb{Z}} \Psi^{j}(x)z^{x} \quad(j=L,0,R).
\end{align*}
Then we have the following lemma.
\begin{lem}
For the three-state Grover walk, we have
\begin{align*}
A_G f_G (z) = 
\begin{bmatrix}
0 \\
0 \\
0
\end{bmatrix},
\end{align*}
where
\begin{align*}
A_G = 
\begin{bmatrix}
\lambda + \dfrac{1}{3z} & -\dfrac{2}{3z} & -\dfrac{2}{3z} 
\\
\\
-\dfrac{2}{3} & \lambda + \dfrac{1}{3} & -\dfrac{2}{3} 
\\
\\
-\dfrac{2z}{3} & -\dfrac{2z}{3} & \lambda + \dfrac{z}{3}
\end{bmatrix}, 
\qquad 
f_G (z) = 
\begin{bmatrix}
f_G ^{L} (z) \\
f_G ^{0} (z) \\ 
f_G ^{R} (z)
\end{bmatrix}.
\end{align*}
\end{lem}
Here we get
\begin{align*}
\det A_G = \frac{\lambda (\lambda -1)}{3z} h_G(z),
\end{align*}
where
\begin{align*}
h_G (z) = z^{2} + \frac{3 \lambda^2 + 4 \lambda + 3}{\lambda} \> z + 1.
\end{align*}
For the following four $\lambda$'s, $h_G(z)$ has double roots.
\begin{align*}
\lambda_{1} = \frac{-1+ 2 \sqrt{2}i}{3}, \quad \lambda_{2} = \frac{-1- 2 \sqrt{2}i}{3}, \quad \lambda_{3} = \lambda_{4} = -1.
\end{align*}
Then we rewrite 
\begin{align*}
\lambda_{+} = \lambda_{1}, \quad \lambda_{-} = \lambda_{2}, \quad \lambda_{\ast} = \lambda_{3} = \lambda_{4}.
\end{align*}
From Eq.(\ref{yokoyamama3}), we have
\begin{align}
\Psi^{0}(x) 
= \frac{2}{1+3 \lambda} \left( \Psi^{L}(x) + \Psi^{R} (x) \right).
\label{eri3}
\end{align}
Combining this with Eqs.(\ref{yokoyama3}) and (\ref{taikan3}) gives
\begin{align}
\lambda \Psi^{L}(x) 
&= A \Psi^{L}(x+1) + B \Psi^{R} (x+1),
\label{takakura3}
\\
\lambda \Psi^{R}(x)
&= B \Psi^{L}(x-1) + A \Psi^{R} (x-1),
\label{ken3}
\end{align}
where 
\begin{align}
A = \frac{1-\lambda}{1+3 \lambda}, \qquad B = \frac{2(1+\lambda)}{1+3 \lambda}.
\label{kensan3}
\end{align}

By Eqs.(\ref{takakura3}),(\ref{ken3}) and (\ref{kensan3}), we see that $\Psi^{L}(x)$ and $\Psi^{R}(x)$ satisfy the following same equation:
\begin{align}
a_{x+2} + \frac{3 \lambda^2 + 4 \lambda + 3}{\lambda} a_{x+1} + a_x = 0,
\label{aida3}
\end{align}
for any $x \in \mathbb{Z}$. We should remark that the characteristic polynomial for difference equation (\ref{aida3}) is equivalent to $h_G(z)=0$.
\par
\
\par\noindent
(i) $\lambda_{\pm} = \frac{-1\pm 2 \sqrt{2}i}{3}$ case. In this case, the solution of $h_G(z)=0$ is $z=-1$ (double roots). So we put
\begin{align}
\Psi^{L}(x) = (-1)^x \Psi^{L}(0), \qquad \Psi^{R}(x) = (-1)^x \Psi^{R}(0). 
\label{mihune3}
\end{align}
From  Eqs.(\ref{takakura3}),(\ref{ken3}) and (\ref{mihune3}), we see that there exists $\psi_0 (\not=0) \in \CM$ such that
\begin{align}
\Psi^{L}(x) = \Psi^{R}(x) = (-1)^x \psi_0. 
\label{tosirou3}
\end{align}
Combining Eqs.(\ref{eri3}) and (\ref{tosirou3}) implies
\begin{align*}
\Psi^{0}(x) = \frac{4 (-1)^x \psi_0}{1 + 3 \lambda_\pm} = \mp \sqrt{2}(-1)^x \psi_0 i. 
%\label{tosirousan3}
\end{align*}
Then we put
\begin{align*}
\Psi^{(\pm)} (x)
= 
\begin{bmatrix}
\Psi^{L}(x) \\
\Psi^{0}(x) \\
\Psi^{R}(x)
\end{bmatrix}
=
(-1)^x \psi_0 
\begin{bmatrix}           
1 \\
\mp \sqrt{2} i \\
1
\end{bmatrix}.
\end{align*}
Therefore 
\begin{align*}
\Psi^{(\pm)} \in {\cal W}^{(\lambda_\pm)}.
\end{align*}
So 
\begin{align*}
\mu^{(\pm)} (x) = c \quad (x \in \ZM),
\end{align*}
where $c = 4 |\psi_0|^2$. Thus $\mu^{(\pm)}$ becomes the stationary measure for the QW and uniform measure with parameter $c = 4 |\psi_0|^2$, that is,
\begin{align*}
\mu^{(\pm)} = \mu^{(c)}_{u, \ZM} \in {\cal M}_s.
\end{align*}
\par
\
\par\noindent
(ii) $\lambda_{\ast} = -1$ case. In this case, the solution of $h_G(z)=0$ is $z=1$ (double roots). So we have
\begin{align}
\Psi^{L}(x) = \Psi^{L}(0), \qquad \Psi^{R}(x) = \Psi^{R}(0). 
\label{yamanba3}
\end{align}
Combining Eqs.(\ref{eri3}) and (\ref{yamanba3}) implies
\begin{align*}
\Psi^{0}(x) = - (\Psi^{L}(0) + \Psi^{R}(0)).
%\label{yamanbasan3}
\end{align*}
Then we have
\begin{align*}
\Psi^{(-1)} (x)
= 
\begin{bmatrix}
\Psi^{L}(x) \\
\Psi^{0}(x) \\
\Psi^{R}(x)
\end{bmatrix}
=
\begin{bmatrix}           
\Psi^{L}(0) \\
- (\Psi^{L}(0) + \Psi^{R}(0)) \\
\Psi^{R}(0)
\end{bmatrix}.
\end{align*}
Therefore 
\begin{align*}
\Psi^{(-1)} \in {\cal W}^{(-1)}.
\end{align*}
So 
\begin{align*}
\mu^{(-1)} (x) = c \quad (x \in \ZM),
\end{align*}
where $c=|\Psi^{L}(0)|^2 + |\Psi^{L}(0) + \Psi^{R}(0)|^2 + |\Psi^{R}(0)|^2.$ Thus $\mu^{(-1)}$ becomes the stationary measure for the QW and uniform measure with parameter $c$; 
\begin{align*}
\mu^{(-1)} = \mu^{(c)}_{u, \ZM} \in {\cal M}_s.
\end{align*}
\par
\
\par\noindent
Finally we consider the following trivial case. 
\par
\
\par\noindent
(iii) $\lambda_{\ast \ast} = 1$ case. In this case, Eqs.(\ref{yokoyama3}), (\ref{yokoyamama3}) and (\ref{taikan3}) implies that there exists $\psi (\not=0) \in \CM$ such that 
\begin{align*}
\Psi^{L}(x) = \Psi^{0}(x) = \Psi^{R}(x) = \psi. 
\end{align*}
Then we put
\begin{align*}
\Psi^{(1)} (x)
= 
\begin{bmatrix}
\Psi^{L}(x) \\
\Psi^{0}(x) \\
\Psi^{R}(x)
\end{bmatrix}
=
\psi
\begin{bmatrix}           
1 \\
1 \\
1
\end{bmatrix}.
\end{align*}
That is, $\Psi^{(1)}$ is the uniform amplitude with parameter $\psi$; 
\begin{align*}
\Psi^{(1)} = \Psi^{(\psi)} _{u,\ZM}. 
\end{align*}
Therefore 
\begin{align*}
\Psi^{(1)} \in {\cal W}^{(1)}.
\end{align*}
So 
\begin{align*}
\mu^{(1)} (x) = 3 |\psi|^2 \quad (x \in \ZM).
\end{align*}
Thus $\mu^{(1)}$ becomes the stationary measure for the QW and uniform measure with parameter $c=3 |\psi|^2$, that is,
\begin{align*}
\mu^{(1)} = \mu^{(3 |\psi|^2)}_{u, \ZM} \in {\cal M}_s.
\end{align*}

In general, the $N$-state Grover walk on $\ZM$ is determined by the $N \times N$ unitary matrix $U^{(G,N)} = [U^{(G,N)}(i,j)]_{1 \le i,j \le N}$. We call $U^{(G,N)}$ $N \times N$ Grover matrix. Here $U^{(G,N)} (i,j)$ is the $(i,j)$ component of $U^{(G,N)}$ given by 
\begin{align*}
U^{(G,N)} (i,i) = \frac{2}{N} -1, \quad U^{(G,N)} (i,j) = \frac{2}{N} \>\> (i \not=j).
\end{align*}
For $k=1,2, \ldots, N$, we put 
\begin{align*}
U^{(G,N)}_k (i,j) = U^{(G,N)} (i,j) \delta_{i,k}.
\end{align*}
Remark that $U^{(G,N)}$ is divided into $\{U^{(G,N)}_k : k=1,2, \ldots ,N \}$, i.e., 
\begin{align*}
U^{(G,N)} = \sum_{k=1}^N U^{(G,N)}_k. 
\end{align*}
For $N=2M+1$ with $M=1,2, \ldots$, $U^{(G,N)}_k$ corresponds to the weight of jump from $x$ to $x - M+ (k-1)$, where $k=1,2, \ldots, N(=2M+1)$. So the range of the jump is $\{x-M, x-M+1, \ldots, x+M-1, x+M\}$. For example, $M=1$ case is the three-state Grover walk. Similarly, for $N=2M$ with $M=1,2, \ldots$, $U^{(G,N)}_k$ corresponds to the weight of jump from $x$ to $x - M + (k-1) \>\> (k=1,2, \ldots, M)$, and from $x$ to $x - M + k \>\> (k=M+1,M+2, \ldots, N=2M)$. So the range of the jump is $\{x-M, x-M+1, \ldots, -1, 1, \ldots , x+M-1, x+M\}$. Then $M=1$ case is the two-state Grover walk considered in Sect. \ref{azero} ($a=0$ case). In a general case also, we have the same argument as both of the $M=1$ case. Let $\Psi^{(1)} = {}^T [\ldots, \psi, \psi, \psi, \ldots]$, that is, its component is always $\psi$. So for the trivial eigenvalue $\lambda_{\ast \ast} = 1$, we have $\Psi^{(1)} \in {\cal W}^{(1)}$. Thus
\begin{align*}
\mu^{(1)} = \mu^{(N |\psi|^2)}_{u, \ZM} \in {\cal M}_s.
\end{align*}

From now on we come back to the three-state case. For the following initial state:
\begin{align*}
\Psi_0 ^{\{0\}}
&= {}^T \left[ \ldots, 
\begin{bmatrix} \Psi^{L} (-2) \\ \Psi^{0} (-2) \\ \Psi^{R} (-2) \end{bmatrix}, 
\begin{bmatrix} \Psi^{L} (-1) \\ \Psi^{0} (-1) \\ \Psi^{R} (-1) \end{bmatrix}, 
\begin{bmatrix} \Psi^{L} (0)  \\ \Psi^{0} (0) \\ \Psi^{R} (0)  \end{bmatrix}, 
\begin{bmatrix} \Psi^{L} (1)  \\ \Psi^{0} (1) \\ \Psi^{R} (1)  \end{bmatrix}, 
\begin{bmatrix} \Psi^{L} (2)  \\ \Psi^{0} (2) \\ \Psi^{R} (2)  \end{bmatrix}, 
\ldots \right],
\\
&= {}^T \left[ \ldots, 
\begin{bmatrix} 0 \\ 0 \\ 0 \end{bmatrix}, 
\begin{bmatrix} 0 \\ 0 \\ 0 \end{bmatrix}, 
\begin{bmatrix} \alpha  \\ \beta \\ \gamma \end{bmatrix}, 
\begin{bmatrix} 0 \\ 0 \\ 0  \end{bmatrix}, 
\begin{bmatrix} 0 \\ 0 \\ 0  \end{bmatrix}, 
\ldots \right],
\end{align*}
where $\alpha, \beta, \gamma \in \CM$ with $|\alpha|^2+|\beta|^2+|\gamma|^2=1$, the following two theorems are shown in Konno \cite{Konno2008b}, which are the generalization of Inui et al. \cite{iks2005}:   
\begin{thm}
\label{konno-aoitori}
\begin{align*}
&
\lim_{n \to \infty} \mu_{n} (x) = \mu_{\infty} (x) = \mu_{\infty} (x;\alpha, \beta, \gamma)
\\
&= 
\left\{
\begin{array}{lc}
\left\{ (3 + \sqrt{6})|2 \alpha + \beta|^2 + (3 - \sqrt{6})|\beta + 2 \gamma|^2 - 2|\alpha + \beta + \gamma|^2 \right\} & \\
\qquad \qquad \qquad \qquad \qquad \qquad \qquad \qquad \qquad \times (49-20 \sqrt{6})^x, & (x \ge 1), \\
& \\
\frac{5 - 2 \sqrt{6}}{2} \left( |2 \alpha + \beta|^2 + |\beta + 2 \gamma|^2 \right), & (x=0), \\
& \\
\left\{ (3 - \sqrt{6})|2 \alpha + \beta|^2 + (3 + \sqrt{6})|\beta + 2 \gamma|^2 - 2|\alpha + \beta + \gamma|^2 \right\} & \\
\qquad \qquad \qquad \qquad \qquad \qquad \qquad \qquad \qquad \times (49-20 \sqrt{6})^{-x}, & (x \le -1), \\
\end{array}
\right. 
\end{align*}
where $49-20 \sqrt{6} = 0.010205 \ldots$, 
\end{thm}
and
\begin{thm}
\label{konno-threestateGroverwalk}
\begin{align*}
{X_n \over n} \quad \Rightarrow \quad Z \qquad (n \to \infty),
\end{align*}
where $Z$ is determined by the following measure: 
\begin{align*}
\mu (dx) 
&= \Delta (\alpha, \beta, \gamma) \> \delta_{0}(dx) + (c_0 + c_1 x + c_2 x^2) \> f_K \left( x ; 1/\sqrt{3} \right) dx, 
\\
&= \Delta (\alpha, \beta, \gamma) \> \delta_{0}(dx) + \frac{ \sqrt{2} (c_0 + c_1 x + c_2 x^2) }{\pi (1-x^2) \sqrt{1-3x^2} } \> I_{(-1/\sqrt{3},1/\sqrt{3})} (x) dx.
\end{align*}
Here $\delta_{0}(dx)$ is the delta measure at the origin and 
\begin{align*}
\Delta(\alpha, \beta, \gamma) 
&= 
\frac{\sqrt{6}-2}{4} \left(|2 \alpha + \beta|^2 + |2 \gamma + \beta|^2 \right)- \frac{5 \sqrt{6}-12}{6} |\alpha + \beta + \gamma|^2,
\\
c_0 
&= \frac{|\alpha + \gamma|^{2}}{2} + |\beta|^{2}, \quad c_1 = - |\alpha - \beta|^{2} + |\gamma - \beta|^{2}, 
\\
c_2 
&= \frac{|\alpha - \gamma|^{2}}{2} - \Re \left( (2\alpha + \beta) (2\overline{\gamma}+\overline{\beta} )\right).
\end{align*}
\end{thm}
Remark that 
\begin{align*}
\Delta(\alpha, \beta, \gamma) = \sum_{x \in  \ZM} \mu_{\infty} (x;\alpha, \beta, \gamma).
\end{align*}

Moreover we obtain
\begin{align*}
{\cal M}_{\infty} ^{\{0\}} 
&= \overline{{\cal M}}_{\infty} ^{\{0\}} 
\\
&= \left\{ \mu_{\infty} (x;\alpha, \beta, \gamma) : \alpha, \beta, \gamma \in \CM \> \hbox{with} \> |\alpha|^2+|\beta|^2 + |\gamma|^2 =1 \> 
\right.
\\
& \left. \qquad \qquad \qquad \qquad \qquad \qquad \qquad \qquad 
\hbox{and} \> |2 \alpha + \beta|^2 + |\beta + 2 \gamma|^2 >0 \right\},
\\ 
{\cal M}^{(w,\{0\})}
&= \left\{ \Delta (\alpha, \beta, \gamma) \> \delta_{0}(dx) + (c_0 + c_1 x + c_2 x^2) \> f_K \left( x ; 1/\sqrt{3} \right) dx 
\right.
\\
& \left. \qquad \qquad \qquad \qquad \qquad \qquad \qquad \qquad : \alpha, \beta, \gamma \in \CM \> \hbox{with} \> |\alpha|^2+|\beta|^2 + |\gamma|^2 =1 \right\}.
\end{align*}
Thus we see that for any $c>0$ and $-1 \le c_1<c_2 \le 1$, 
\begin{align*}
\mu_{u,\ZM} ^{(c)} \not\in {\cal M}_{\infty} ^{\{0\}} = \overline{{\cal M}}_{\infty} ^{\{0\}}, \quad \mu_{u,(c_1,c_2)} ^{(c)} \not\in  {\cal M}^{(w,\{0\})}.
\end{align*}

\section{Summary \label{sum}}
In this paper we proved 
\begin{thm}
Let $U$ be $2 \times 2$ unitary matrix or $N \times N$ Grover matrix $(N \ge 3)$. For the QW on $\ZM$ determined by $U$, we have 
\begin{align*}
\mu_{u, \ZM}^{(c)} \in {\cal M}_s (U),
\end{align*}
where $c>0$. Here $\mu_{u, \ZM}^{(c)}$ is the uniform measure with parameter $c$, i.e., $\mu_{u, \ZM}^{(c)}(x) = c$ for any $x \in \ZM$ and ${\cal M}_s (U)$ is the set of stationary measures of the QW defined by $U$.
\end{thm}
Moreover we presented a method for producing the uniform probability measure in Sect. \ref{intro}. As a future work, it would be interesting to investigate the relation between stationary measure, time-averaged limit measure, and weak limit measure for the QW in more general setting.

\par
\
\par\noindent
{\bf Acknowledgment.} This work was partially supported by the Grant-in-Aid for Scientific Research (C) of Japan Society for the Promotion of Science (Grant No.24540116).

\par
\
\par

\begin{small}
\bibliographystyle{jplain}

\end{small}

\end{document}